\documentstyle[psfig]{mn}
\newcommand{\Msun}{M_{\odot}}
\newcommand{\Zsun}{Z_{\odot}}

\newcommand{\average}[1]{\langle\rm{#1}\rangle_{\rm V}}

\newcommand{\etal}{{et~al.}}

\newcommand{\Ae}{$\alpha$-enhancement}

\title{Abundance ratios in hierarchical galaxy formation}

\author[D.~Thomas]
{D.~Thomas\\
Universit\"ats-Sternwarte M\"unchen, Scheinerstr.~1, D-81679 M\"unchen,
Germany}

\begin{document}

\bibliographystyle{mnras}

\maketitle
\begin{abstract}
The chemical enrichment and stellar abundance ratios of galaxies which form
in a hierarchical clustering scheme are calculated. For this purpose I adopt
the star formation histories (SFH) as they are delivered by semi-analytic
models in Kauffmann \shortcite{K96}. It turns out that the average SFH of
cluster ellipticals does not yield globally $\alpha$-enhanced stellar
populations. The star burst that occurs when the elliptical forms in the
major merger plays therefore a crucial role in producing \Ae. Only under the
assumption that the IMF is significantly flattened with respect to the
Salpeter value during the burst, a Mg/Fe overabundant population can be
obtained. In particular for the interpretations of radial gradients in
metallicity and \Ae, the mixing of global and burst populations are of great
importance. The model predicts bright field galaxies to be less
$\alpha$-enhanced than their counterparts in clusters.
\end{abstract}

%================= I N T R O D U C T I O N =================================

\section{Introduction}
\label{intro}
Simulations of hierarchical galaxy formation in a CDM universe (Kauffmann,
White \& Guiderdoni 1993; Cole \etal\ 1994)\nocite{KWG93,Cetal94} are based
on semi-analytic models in the framework of Press-Schechter theory. They aim
to describe the formation of galaxies in a cosmological context, and
therefore are designed to match a number of constraints like $B$ and $K$
luminosity functions, the Tully-Fisher relation, redshift distribution, and
slope and scatter of the colour-magnitude relation (CMR). Since in a
bottom-up scenario more massive objects form later, and are therefore
younger and bluer -- in contrast to the observed CMR -- the original works
were substantially suffering from the problem of creating luminous red
elliptical galaxies \cite{KWG93,Letal93,Cetal94}. It is also not a priori
clear if these models which yield strong evolution at intermediate redshift
(Kauffmann, Charlot \& White 1996)\nocite{KCW96} would be able to reproduce
the small scatter of the CMR and the Mg-$\sigma$ relation. On the other
hand, these constraints can be easily explained by the classical single
burst picture for elliptical galaxy formation, assuming passive evolution
after a short formation epoch at high redshift, provided that subsequent
stellar merging is negligible (Bower, Kodama \& Terlevich
1998)\nocite{BKT98}. However, the more recent models by Kauffmann (1996,
hereafter K96)\nocite{K96} and Baugh, Cole \& Frenk \shortcite{BCF96}
reproduce the above relations with a scatter which is in remarkably good
agreement with observational data (Bower, Lucey \& Ellis 1992; Bender,
Burstein \& Faber 1993; J{\o}rgensen, Franx \& Kj{\ae}rgaard
1996)\nocite{BLE92,BBF93,JFK96}. The correct slope of the CMR can be
obtained in a hierarchical scheme, if metal enrichment and
metallicity-dependent population synthesis models are taken into account
(Kauffmann \& Charlot 1998; Cole et al., in preparation)\nocite{KC98}. In
such models, the slope of the CMR is only driven by metallicity, as in the
classical models by Arimoto \& Yoshii
\shortcite{AY87}. Note however, that in the framework of the inverse wind
models \cite{M94} the CMR slope could be in principle produced from a
combination of both age and metallicity, since in these models more massive
galaxies are assumed to be older.

In this paper I aim to discuss how far hierarchical formation models are
able to accomplish a further important constraint, namely the formation of
$\alpha$-enhanced stellar populations hosted by luminous ellipticals
(Peletier 1989; Worthey, Faber \& Gonz\'{a}lez 1992; Davies, Sadler \&
Peletier 1993)\nocite{P89,WFG92,DSP93}. Models of chemical evolution show
that this constraint can be matched by the single collapse model being
characterised by short star formation time-scales of the order $10^8-10^9$
yr (e.g.\ Matteucci 1994; Thomas, Greggio \& Bender 1999; Jimenez \etal\
1999)\nocite{M94,TGB99,Jetal99}, and it has been questioned by Bender
\shortcite{Be97} whether the observed [Mg/Fe] overabundance is compatible
with hierarchical models. However, abundance ratios and the enrichment of
$\alpha$-elements have not been investigated so far in hierarchical models.
For this purpose, I consider typical star formation histories provided by
semi-analytic models for hierarchical galaxy formation (K96) and explore the
resulting abundance ratios of magnesium and iron.

The paper is organised as follows. In Section~\ref{model} the model of
chemical enrichment is briefly described. Average and bursty star formation
histories are then analysed in Sections~\ref{average}, \ref{bursty}
and~\ref{mixtures}. The results are discussed and summarised in
Sections~\ref{discussion} and~\ref{conclusions}.

%=================== S E C T I O N 2 ======================================

\section{The model}
\label{model}
The idea is to follow the chemical evolution of the galaxy describing the
global chemical properties of the final object and its progenitors as a
whole during the merging history. In the hierarchical clustering scheme,
structures are subsequently build up starting from small disc-like objects.
An elliptical is formed when two disc galaxies of comparable mass merge,
which is called the `major merger'. Before this event many `minor mergers'
between a central galaxy and its sattelite systems happen. It is important
to emphasise, that the bulk of stars, namely $70-90$ per cent, forms at
modest rates in the progenitor disc-galaxies {\em before} this `major
merger' event. In the star burst ignited by the latter, up to 30 per cent of
the total stellar mass of the elliptical is created. In K96 many Monte Carlo
simulations are performed each corresponding to one individual (final)
galaxy. The SFH which is specified in K96 and adopted for the present
analysis is an {\em average} about all these realizations.

\subsection{Star formation rates}
The SFH is characterised by the age distribution of the stellar populations
weighted in the $V$-band light. In other words, the fractional contributions
$L_{V}^{\rm SP}$ by stellar populations to the total $V$-light as a function
of their ages $t$ are specified. Equation~\ref{trans} shows how this
translates into a star formation {\em rate} $\psi$:
\begin{equation}
L_{V}^{\rm SP}(t_0,t_1) =
\frac{\int_{t_{\rm 0}}^{t_{\rm 1}} \psi(t)\; L_{V}^{\rm SSP}(t)\: dt}
{\int_{0}^{t_{\rm univ}} \psi(t)\; L_{V}^{\rm SSP}(t)\: dt}\ .
\label{trans}
\end{equation}
Here, the interval $[t_0,t_1]$ denotes the age bin of the population,
$t_{\rm univ}$ is the assumed age of the universe, i.e.\
$t_{\rm\-univ}\approx 13$~Gyr for the cosmology adopted in K96
($\Omega_m=1,\-\Omega_\Lambda=0, h=0.5$). The $V$-light of a simple stellar
population $L_{V}^{\rm SSP}(t)$ as a function of its age and metallicity is
taken from Worthey \shortcite{Wo94}. In this paper, the star formation rate
$\psi$ as a function of time is chosen such that a set of stellar
populations of various ages and metallicities is constructed which exactly
covers the age distribution of the K96 models.

\subsection{Chemical enrichment}
The chemical evolution is calculated by solving the usual set of
differential equations (e.g.\ Tinsley 1980)\nocite{T80}. The enrichment
process of the elements hydrogen, magnesium, iron, and total metallicity is
computed, no instantaneous recycling is assumed. In particular, the delayed
enrichment by Type~Ia supernovae is taken into account following the model
by Greggio \& Renzini \shortcite{GR83}. The inclusion of Type Ia is crucial
for interpreting Mg/Fe ratios, since SNe~Ia substantially contribute to the
enrichment of iron. The evolution code is calibrated on the chemical
evolution in the solar neighbourhood (Thomas, Greggio \& Bender
1998)\nocite{TGB98}. The ratio of SNe~Ia to SNe~II is chosen such that
observational constraints like supernova rates, the age-metallicity
relation and the trend of Mg/Fe as a function of Fe/H in the solar
neighbourhood are reproduced. The SN~II nucleosynthesis prescription is
adopted from Thielemann, Nomoto \& Hashimoto \shortcite{TNH96} and Nomoto
\etal\ \shortcite{Netal97}, because the
stellar yields from Woosley \& Weaver \shortcite{WW95} are unable to account
for the [Mg/Fe] overabundance in the solar neighbourhood \cite{TGB98}.

The IMF is truncated at $0.1~\Msun$ and $40~\Msun$. The slope above
$1~\Msun$ is treated as a parameter, below this threshold a flattening
according to recent HST measurements (Gould, Bahcall \& Flynn
1997)\nocite{GBF97} is assumed. This flattening at the low-mass end does not
significantly affect $\alpha$/Fe ratios. It should be noted that stars more
massive than $40~\Msun$ are expected to play a minor role in the
enrichment process, because of their small number and the fall-back
effect \cite{WW95}. Moreover, the mean [Mg/Fe] overabundance in the
metal-poor halo stars of our Galaxy is well reproduced for an IMF with
the above truncation and Salpeter slope $x=1.35$ \cite{S55}, if
Thielemann \etal\ nucleosynthesis is adopted \cite{TGB98}. The
$V$-luminosity averaged abundances in the stars are computed by using
SSP models from Worthey \shortcite{Wo94}. For details I refer the
reader to Thomas \etal\ (1998, 1999)\nocite{TGB98,TGB99}.

\smallskip
Finally, two simplifications of the present model should be mentioned:
\begin{enumerate}
\item A one-phase interstellar medium (ISM) is assumed. Kauffmann \& Charlot 
\shortcite{KC98}, instead, distinguish between cold and hot gas
allowing mass transfer among these two components. Since stars form out of
cold gas, this process basically controls the feed-back mechanism of star
formation. The resulting effects on the star formation rate, however, are
already covered, since I directly adopt the SFH from K96. The influence on
the abundance {\em ratio} of Mg and Fe is expected to be negligible, as long
as these elements do not mix on significantly different time-scales.
\item The final galaxy is treated as the whole single-unit from the
beginning of its evolution, the economics of the individual progenitors are
not followed separately as in K96. Hence, the simulations do not directly
include galactic mass loss taking place during disc galaxy evolution (K96)
as well as mass transfer between the progenitor systems. The fraction of gas
converted to stars is adjusted to global metallicities typically observed in
the respective galaxy type. The impact on the {\em global} properties of the
{\em final} galaxy, however, is expected to be small. In particular,
abundance {\em ratios} of non-primordial elements like Mg and Fe are not
significantly affected by mass loss and gas transfer, unless one allows for
selective mechanisms. However, in the framework of the present analysis,
this option shall only be discussed in a qualitative fashion.
\end{enumerate}

%===========================  S E C T I O N 3 =============================

\section{Global populations}
\label{average}
\begin{figure*}
\begin{minipage}{0.49\linewidth}
\psfig{figure=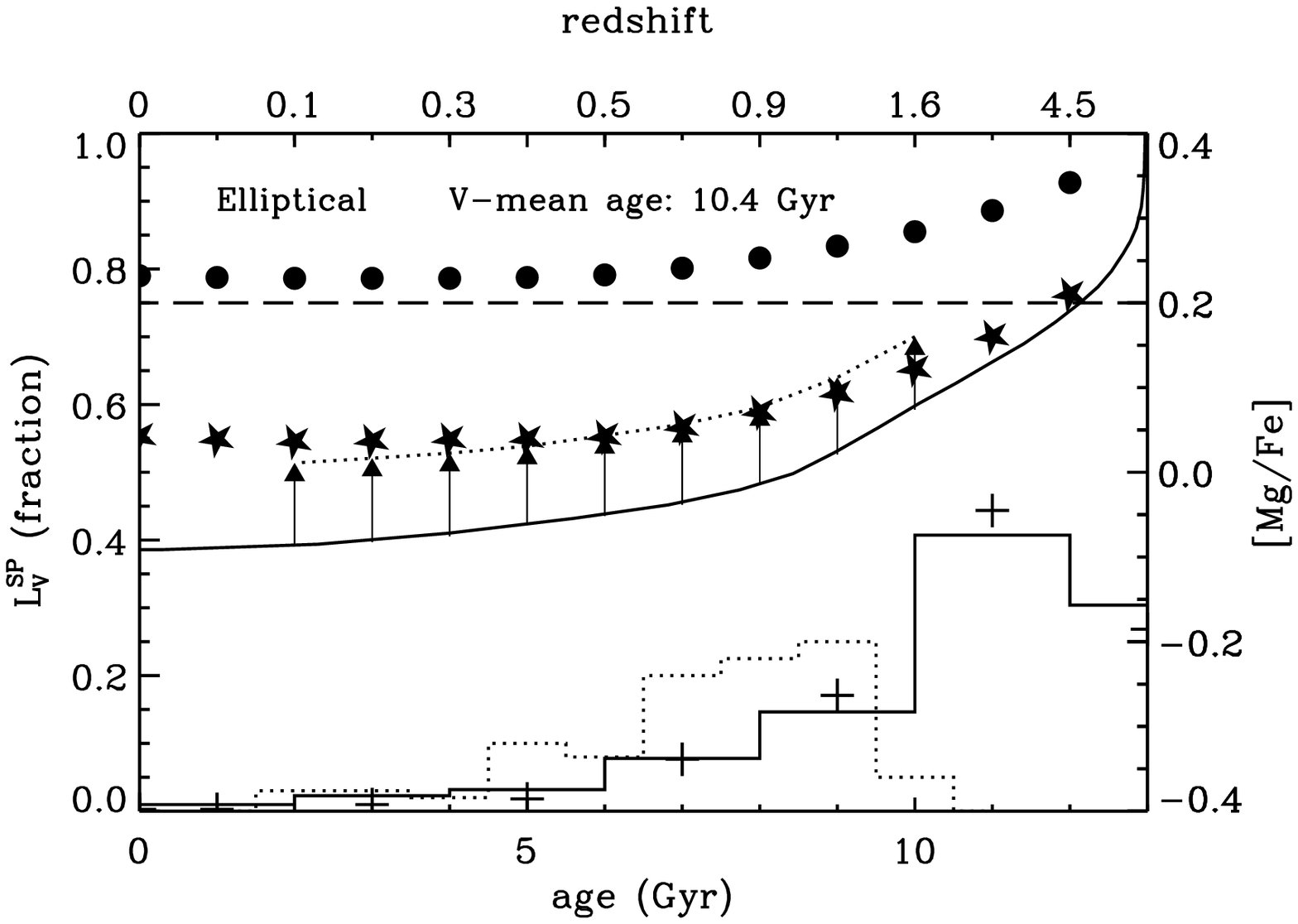,width=\linewidth}
\end{minipage}
\begin{minipage}{0.49\linewidth}
\psfig{figure=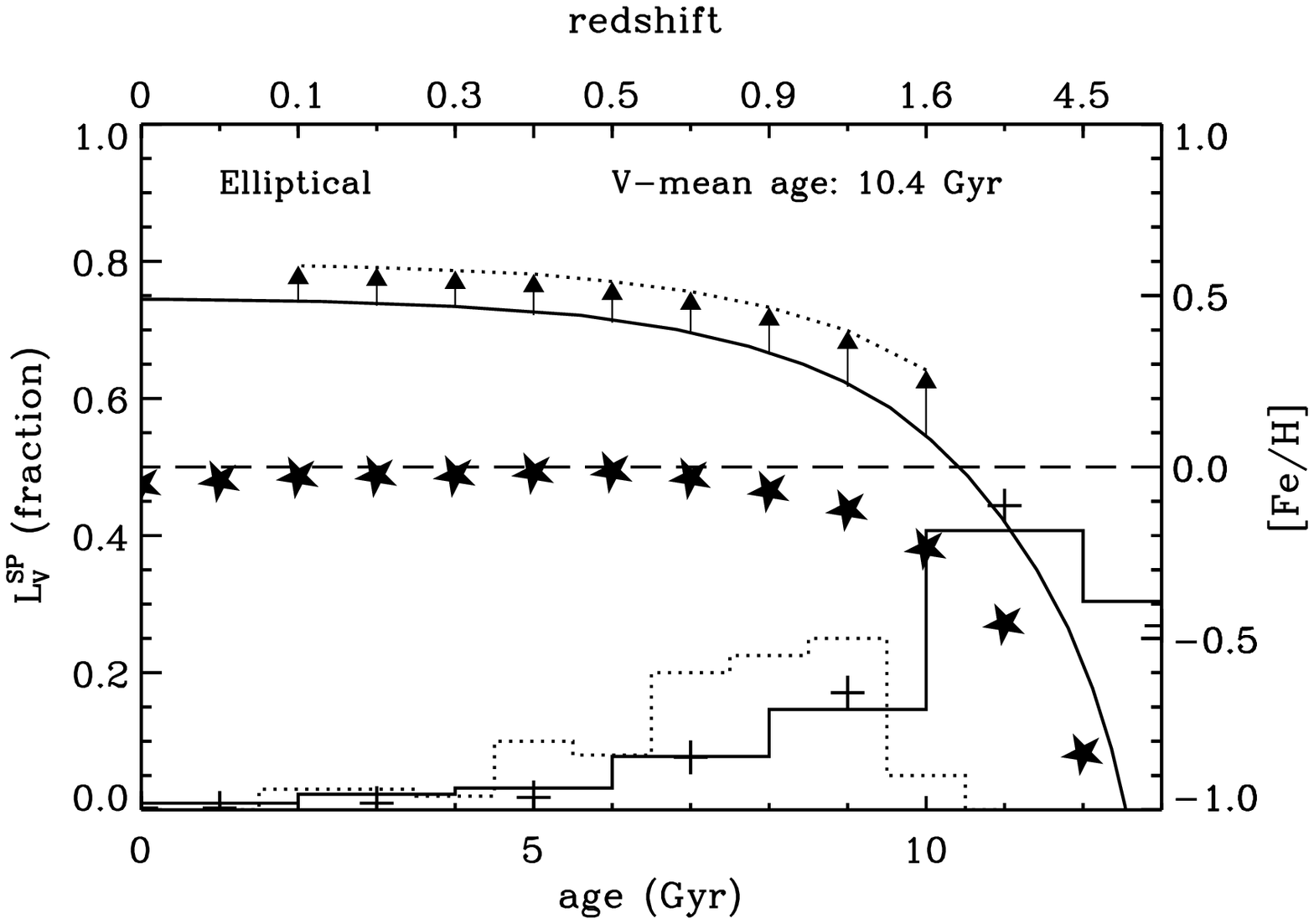,width=\linewidth}
\end{minipage}
\caption{The star formation history of an average cluster elliptical. The
lower and upper x-axis specify age (look-back time) and redshift
($\Omega_m=1,\-\Omega_\Lambda=0$), respectively. The left y-axis denote a
`fraction'. The histogram shows the fractional contribution to the total
light in the $V$-band from each stellar population of a specific age (see
eqn.~\ref{trans}). The plus-signs denote the respective contribution to the
total mass of the object. The scales for the underlying abundance ratios
Mg/Fe (left-hand panel) and Fe/H (right-hand panel) are given by the
right-hand y-axis: interstellar medium abundance (solid line), $V$-average
in the composite stellar population (star symbols), the $V$-average in the
composite stellar population for a flat IMF ($x=0.8$, filled circles). The
dotted histogram indicates the fractional number distribution of the major
merger epochs of elliptical formation as they come out in the individual
Monte Carlo realizations (K96). The arrows to the dotted line at the various
merger epochs give the stellar abundance ratios of the newly created burst
population formed in the major merger (see text). The horizontal dashed
lines mark the observational constraints for \Ae\ and metallicity of
elliptical galaxies, respectively.}
\label{fig1}
\end{figure*}

\begin{figure*}
\begin{minipage}{0.49\linewidth}
\psfig{figure=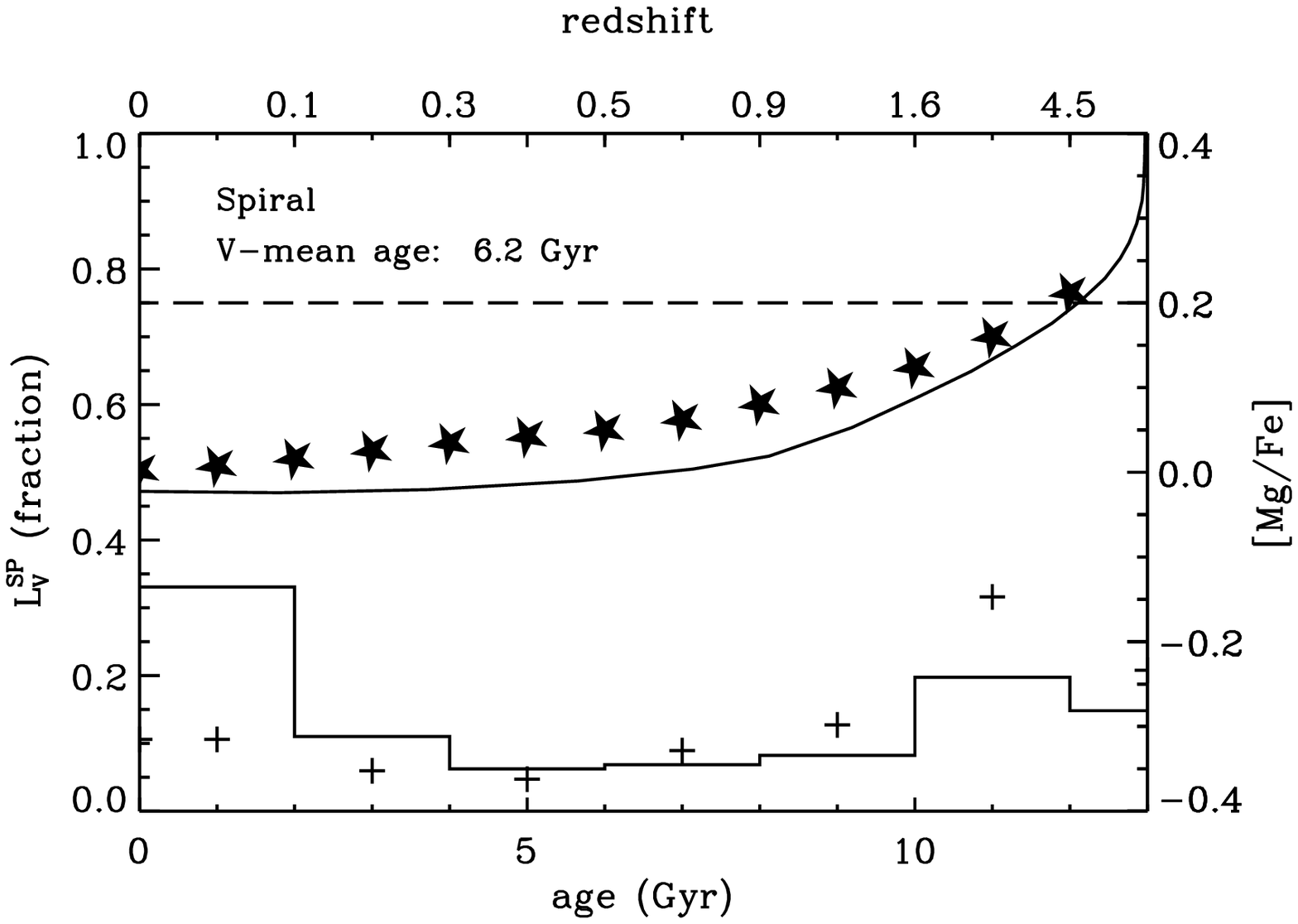,width=\linewidth}
\end{minipage}
\begin{minipage}{0.49\linewidth}
\psfig{figure=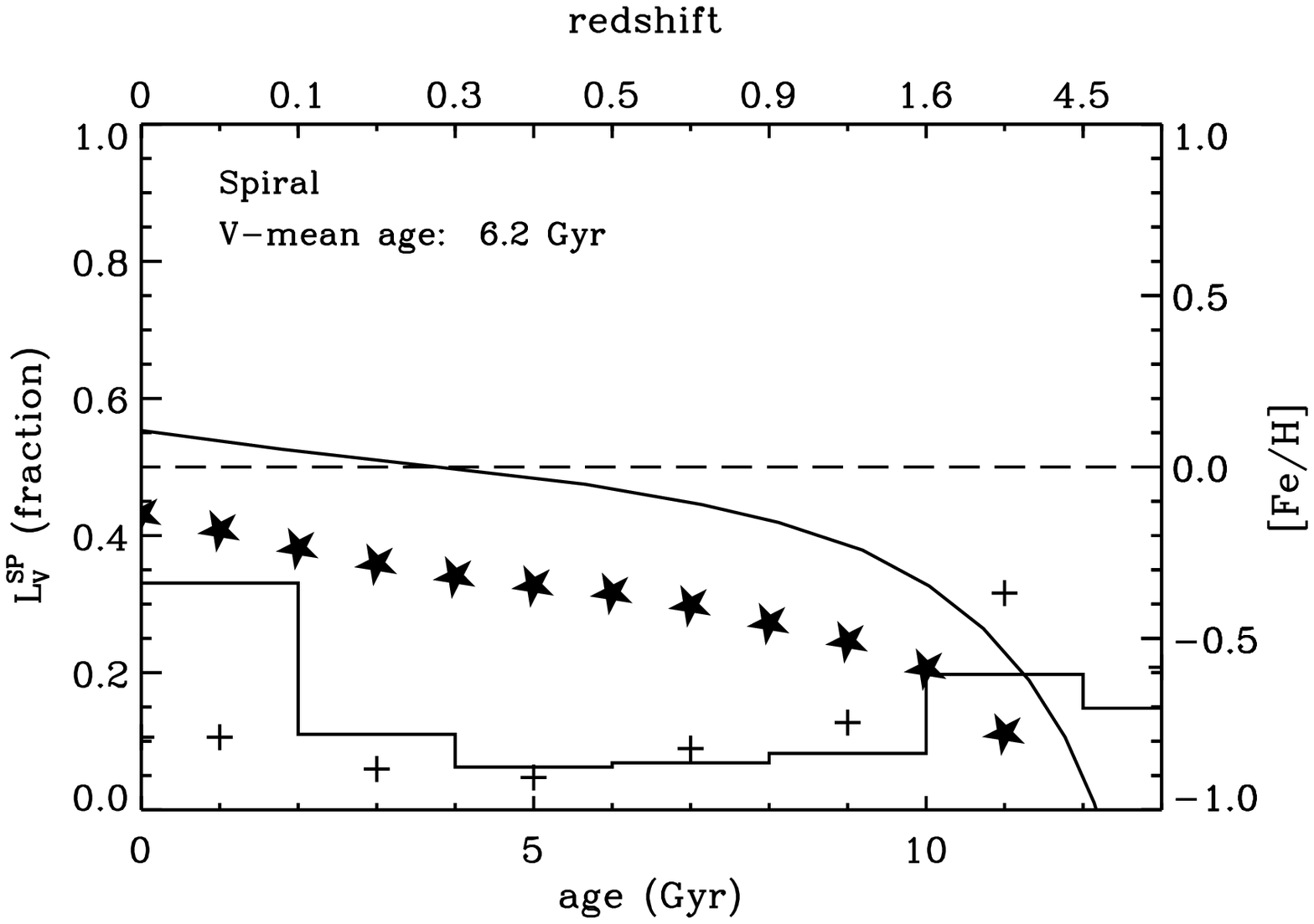,width=\linewidth}
\end{minipage}
\caption{The star formation history of an average spiral. Axis and symbols are
explained in Fig.~\ref{fig1}.}
\label{fig1b}
\end{figure*}

Figs.~\ref{fig1} and~\ref{fig1b} show the global average SFHs as they are
constrained for cluster elliptical and spiral galaxies in the K96 model. The
lower x-axis denote ages (look-back time), the evolutionary direction of
time therefore goes from the right to the left. The evolution with redshift
for an $\Omega_m=1$, $\Omega_\Lambda=0$ cosmology (adopted in K96) is given
by the upper x-axis. The histogram shows the quantity
$L_{V}^{\rm\-SP}(t_0,t_1)$ as it is explained in Section~\ref{model}. In
the elliptical case (Fig.~\ref{fig1}), more than 70 per cent of the light in
the $V$-band comes from stars which have formed in the first 3 Gyr at
redshift $z\ga 2$. The SFH of the spiral (Fig.~\ref{fig1b}), instead, is
much more extended towards lower redshift. With a $V$-mean age of 6.2 Gyr,
the latter thus leads to significantly younger populations than the SFH of
the average elliptical (10.4 Gyr). The plus signs in Figs.~\ref{fig1}
and~\ref{fig1b} denote the fractional contributions from the various
populations to the total mass of the object. Owing to their stellar
remnants, old populations contribute more to the total mass than to the
light of the object. As a consequence, in the optical band the weight of
young (old) populations is larger (smaller) with respect to their actual
mass. This pattern is particularly prominent in the young and old spiral
populations.

The abundance ratios of Mg/Fe (left-hand panels) and Fe/H (right-hand
panels) as they result from the chemical evolution model are shown with the
scale on the right y-axis, respectively. The chemical evolution of the ISM
is denoted by the solid line, the filled star symbols show the
$V$-luminosity weighted average abundance ratios of the respective composite
stellar population of the galaxy as a function of age (hence formation
redshift). The filled circles in the left-hand panel of Fig.~\ref{fig1}
represent the stellar mean Mg/Fe ratio if a global flattening of the IMF
with $x=0.8$ is assumed. Otherwise, Salpeter slope above $1~\Msun$
($x=1.35$) is chosen.

In general, the mean abundances in the stellar populations differ from those
in the ISM. Metallicity (Fe/H) is always higher, and Mg/Fe always lower in
the ISM than in the $V$-average of the stars, because the stars archive the
abundance patterns of early epochs. Since, in the case of the elliptical,
star formation is much more skewed towards early times, this discrepancy is
more significant in such objects. The gas fraction which is ejected from the
galaxy is chosen such that the {\em global} stellar populations of the
elliptical have solar metallicity at low redshift (Fig.~\ref{fig1}). This
leads to an ejection of roughly 30 per cent of the baryonic mass of the
galaxy during its evolution. In the case of the spiral (Milky Way), instead,
55 per cent of the gas are assumed to be ejected in order to yield solar
metallicity in the ISM after 9 Gyr, roughly when the Sun was born
(Fig.~\ref{fig1b}).

As a consequence, the metallicity of the ISM and the stars is higher in the
elliptical than in the spiral. The Mg/Fe ratios, instead, depend mainly on
the shape of the star formation rate as a function of time. Since there is
no significant star formation at late times in the elliptical,
$N_{\rm\-SNII}/N_{\rm\-SNIa}$ is 20 times lower than in the spiral galaxy,
which leads to a lower Mg/Fe in the ISM by roughly 0.1 dex. The stars,
instead, store the chemical abundances of the early epochs,
$\average{[Mg/Fe]}$ is 0.04 dex higher in the elliptical galaxy than in the
spiral, which is in accordance with its older mean age. Still, the stellar
populations in both galaxy types exhibit roughly solar Mg/Fe ratios at low
redshift, in particular $\average{[Mg/Fe]}\approx\-0.04$ dex for the
elliptical. In spite of the negligible contribution of the late star
formation to the mean age of the galaxy, the mean [Mg/Fe] ratio is driven
significantly below $0.2$~dex (Fig.~\ref{fig1}). This result clearly stands
in disagreement to the observational indications that (bright) elliptical
galaxies host stellar populations which are $\alpha$-enhanced by at least
$0.2-0.3$ dex \cite{P89,WFG92,DSP93}.

Fig.~\ref{fig1} demonstrates that with Salpeter IMF a super-solar Mg/Fe
ratio is only obtained at the very early stages of the evolution, namely in
the first Gyr. A star formation mode which is more skewed towards these
early formation ages would lead to older ages and a higher degree of \Ae.
One may argue that the average SFH considered here does not apply to {\em
bright} ellipticals in which the observed \Ae\ is most significant. However,
in the K96 simulations brighter objects are on average younger \cite{KC98},
and do not experience SFHs that are more skewed towards high redshift than
the present example. They are therefore unlikely to exhibit higher Mg/Fe
ratios than calculated here.

The star formation time-scales that are required to obtain
[$\alpha$/Fe]$\approx\-0.2$ for different IMF slopes are discussed in Thomas
\etal\ \shortcite{TGB99}. The filled circles in the upper-left panel of
Fig.~\ref{fig1} demonstrate that the shallower IMF with $x=0.8$ can
reconcile the extended SFH with \Ae. However, particularly in the
hierarchical picture in which the formation of different galaxy types can be
traced back to the same (or similar) building blocks at early epochs, a
flattening of the IMF restricted to elliptical galaxies does not seem to
provide a compelling solution.

From Fig.~\ref{fig1} one can also understand that only galaxy formation
scenarios in which star formation {\em terminates} after $1-2$~Gyr lead to
$\average{[Mg/Fe]}\ga 0.2$ dex. The classical monolithic collapse provides
such a star formation mode. In the hierarchical scheme, instead, (low) star
formation at later stages is unavoidable, which directly leads to lower
Mg/Fe ratios in the stars that form out of the highly SNIa enriched ISM at
lower redshift.

%=================================  S E C T I O N 4 ========================

\section{Burst populations}
\label{bursty}
\begin{figure*}
\begin{minipage}{0.49\linewidth}
\psfig{figure=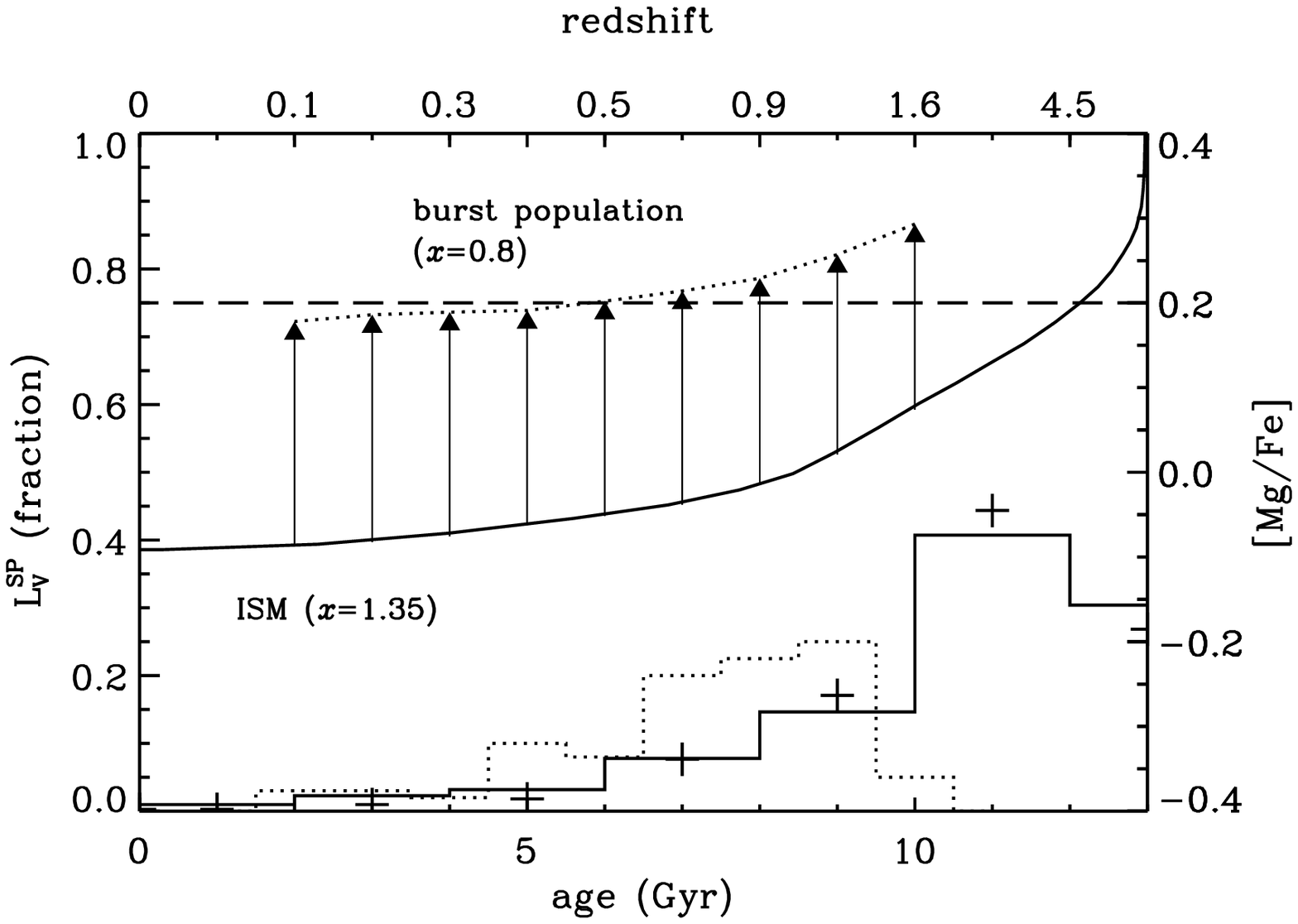,width=\linewidth}
\end{minipage}
\begin{minipage}{0.49\linewidth}
\psfig{figure=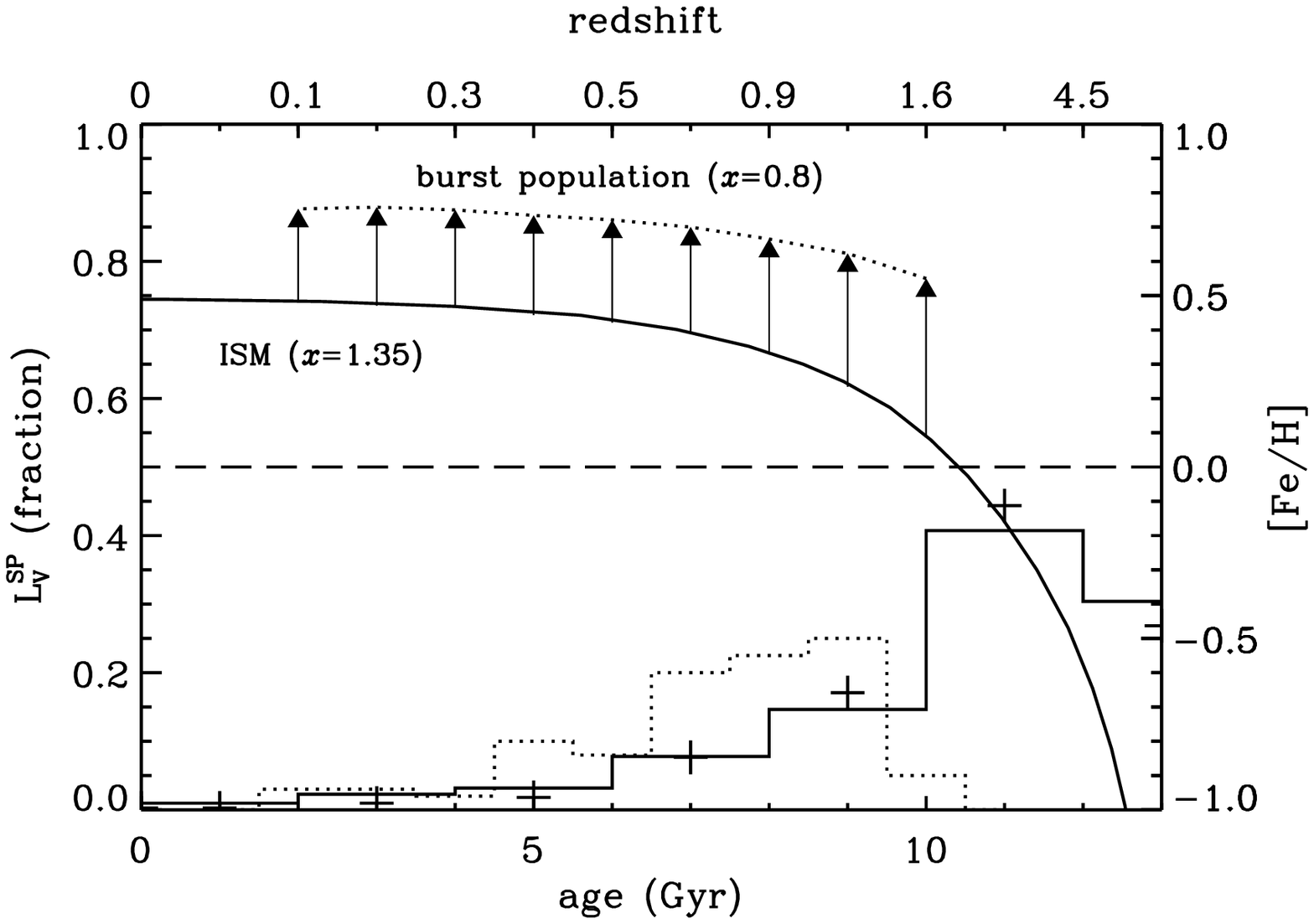,width=\linewidth}
\end{minipage}
\caption{Abundance ratios for the average cluster elliptical. Axis and
symbols are explained in Fig.~\ref{fig1}. A flattening of the IMF ($x=0.8$)
during the major merger burst is assumed. For the evolution of the global
interstellar medium (solid line), instead, Salpeter IMF ($x=1.35$) is
adopted.}
\label{fig2}
\end{figure*}
The SFHs considered here apply to the {\em global} properties of the galaxy
populations. However, in the picture of hierarchical clustering star bursts
that are induced by mergers play an important role for the interpretation of
(particularly central) abundance features. The star burst during the major
merger forms $10-30$ per cent of the final stellar mass, which is likely to
dominate the galaxy core, since the gas is driven to the central regions of
the merger remnant \cite{BH96}. In this section I shall investigate the
abundance patterns of the population that results from a $0.1$~Gyr star
burst triggered by the major merger which forms the final elliptical.

\subsection{Universal IMF}
In Fig.~\ref{fig1} the dotted histogram denotes the distribution of these
major merger events among the elliptical galaxy population. At the given
epoch when the star burst occurs, the new stellar population forms out of
the ISM whose abundances are shown by the solid line in Fig.~\ref{fig1}.
Owing to the short duration of the star burst, the mean Mg/Fe ratios in the
newly created stars are higher than the initial values in the ISM as
indicated by the arrows and the dotted line (left-hand panel of
Fig.~\ref{fig1}). The slope of the IMF is assumed Salpeter also during the
burst. It turns out that, due to the extremely low Mg/Fe ratios and the high
metallicities in the ISM at the merger epoch, such star bursts are
inappropriate to raise Mg/Fe up to a level consistent with observational
values. The resulting Mg/Fe does not differ from the mean average in the
global stellar populations.

The metallicity Fe/H, instead, increases to even higher values between 0.3
and 0.5 dex, depending on the burst epoch (upper-right panel of
Fig.~\ref{fig1}). Under the assumption that the entire central population
forms in the major burst the model leads to radial metallicity gradients of
$0.3-0.5$ dex from the inner to the outer parts of the galaxy, in
contradiction to observational measurements of $\sim 0.2$~dex per decade
\cite{DSP93,Ku98}. A mixture of $3/4-1/3$ between burst and global population in
the galaxy nucleus is required to smooth the gradient down to the observed
value.

\smallskip
The results shown in Fig.~\ref{fig1} unfold the principle dilemma of the
hierarchical picture: the initial conditions in the ISM at the burst epoch,
namely super-solar metallicity and sub-solar Mg/Fe, are highly unfavourable
in producing $\alpha$-enhanced populations. As shown in Thomas \etal\
\shortcite{TGB99}, this incapability of late merger events is independent
of the burst time-scale and SN~Ia rates during the burst. The most promising
way out to reconcile an intermediate or late merger with \Ae\ is to assume a
flattening of the IMF during the star burst.

\subsection{Variable IMF}
The arrows and dotted lines in Fig.~\ref{fig2} show the abundance patterns
(left-hand panel: Mg/Fe, right-hand panel: Fe/H) of the burst population if
a shallow IMF ($x=0.8$) {\em during the burst} is assumed. The chemical
evolution of the global ISM is based on Salpeter IMF ($x=1.35$) as before.
It turns out that, with a significant flattening of the IMF, the major
merger burst produces a metal-rich $\alpha$-enhanced stellar population. As
demonstrated in the left-hand panel of Fig.~\ref{fig2}, the bulk of
ellipticals experience their major merger at look-back times of $7-9$~Gyr
($z\approx 0.7-1.5$) and therefore exhibit significant \Ae\ in the burst
population.

\section{Global--burst mixtures}
\label{mixtures}
The question is how do the global and the burst populations mix? The value
$x=0.8$ represents the minimum flattening required for the case that the
pure burst population is observed, i.e.\ in the centre of the galaxy. Apart
from the fact that no mixing of the two population types seems unlikely, it
would produce metallicity gradients much too steep. Table~\ref{tab} gives
the average abundance ratios for different mixtures of the global (IMF slope
$x=1.35$) and the 8~Gyr-burst populations ($x=0.8$). [Fe/H]' denotes the
metallicity if the efficiency of the star burst is reduced from 99 to 50 per
cent. Mg/Fe is not significantly affected \cite{TGB99}.
\begin{table}
\caption{The average abundance ratios for different mixtures in $V$-light
of the global (IMF slope $x=1.35$) and burst ($x=0.8$) populations. [Fe/H]'
denotes the metallicity if the efficiency of the star burst is reduced from
99 to 50 per cent. The burst epoch at 8~Gyr is chosen.}
\begin{tabular}{cccccccc}
\hline
global & burst & [Mg/Fe] & [Fe/H]  & [Fe/H]'   \\\\
 1.0   &  0.0  &   0.04  &  0.00   &  0.00     \\
 0.9   &  0.1  &   0.06	 &  0.10   &  0.05     \\
 0.7   &  0.3  &   0.11	 &  0.30   &  0.18     \\
 0.5   &  0.5  &   0.15	 &  0.44   &  0.29     \\
 0.3   &  0.7  &   0.18	 &  0.54   &  0.37     \\
 0.1   &  0.9  &   0.21	 &  0.63   &  0.45     \\
 0.0   &  1.0  &   0.23  &  0.66   &  0.50     \\
\hline
\end{tabular}
\label{tab}
\end{table}

If the burst population mixes homogeneously with the global population by
providing 30 per cent of the total $V$-light, the resulting [Mg/Fe] is
reduced to 0.11~dex (row~3 in Table~\ref{tab}). For such proportions, the
extreme flattening of $x=0$ would be required to obtain a significantly
$\alpha$-enhanced population of [Mg/Fe]=0.2~dex. More likely, however, is
that the burst populations are more prominent in the inner parts due to
dissipation processes in the burst gas \cite{BH96}. Table~\ref{tab} can then
be interpreted in terms of increasing galaxy radius from bottom to top.
Abundance gradients result according to the global-burst mixtures assumed at
different radii of the object.

I shall briefly discuss the following model, in which I roughly devide the
galaxy in three zones separated by the inner radius $r_i$ and the effective
radius $r_e$. The resulting stellar abundance ratios in each zone for the
respective different global-burst mixtures are shown in Table~\ref{tab2}.
Column~2 gives the $V$-light fraction of each zone, namely 5 per cent from
the most inner part, and half of the light within the effectice radius
$r_e$. Altogether, the fractional contributions are chosen such that the
burst population contributes 30 per cent of the total $V$-light of the
galaxy. The star formation efficiency of the burst population is assumed to
be 50 per cent, in order to avoid metallicities that exceed observational
determinations. In Table~\ref{tab2}, the quota of the burst population is
steadily decreasing towards the outer zones of the galaxy, which leads to
gradients such that both Mg/Fe and Fe/H are decreasing with increasing
radius.

\begin{table}
\caption{Abundance ratios in the three zones separated by $r_i$ and $r_e$.
The first column gives the fraction of total light provided by the
respective zone, cols.~2 and~3 specify the global-burst mixtures within each
zone. [Fe/H]' is explained in the caption of Table~1.}
\begin{tabular}{cccccc}
\hline
zone           & light & global & burst & [Mg/Fe] & [Fe/H]' \\
$0<r<r_i$      & 0.05  &  0.1   &  0.9  &   0.21  &  0.45   \\
$r_i<r<r_e$    & 0.45  &  0.5   &  0.5  &   0.15  &  0.29   \\
$r_e<r<\infty$ & 0.50  &  0.9   &  0.1  &   0.06  &  0.05   \\
\hline
\end{tabular}
\label{tab2}
\end{table}

The galaxy nucleus is most metal-rich and significantly $\alpha$-enhanced,
the gradients of both Mg/Fe and Fe/H within $r_e$, however, are rather
shallow. Thus this model predicts the Mg/Fe overabundance not to decline
remarkably out to $r_e$, in accordance with Worthey \etal\ \shortcite{WFG92}
and Davies \etal\ \shortcite{DSP93}. It is interesting to mention that the
tendancy of a slight decrease in \Ae\ is found in the Coma cluster
sample analysed by Mehlert~\etal\ (in preparation). The gradient in
metallicity as given in Table~\ref{tab2} is also consistent with estimates
by Davies \etal\ \shortcite{DSP93} and Kuntschner \shortcite{Ku98}.

Finally it should be emphasized, that reasonable metallicities are only
achieved if star formation during the burst is assumed to be efficient by
only 50 per cent. Table~\ref{tab} shows that otherwise metallicities above
$3~\Zsun$ would be obtained in the centre, which exceeds observational
determinations \cite{KD98}.

%========================== D I S C U S S I O N ==========================

\section{Discussion}
\label{discussion}
It should again be emphasized that the SFHs discussed here are averages
over many Monte Carlo realizations, individual galaxies are therefore not
considered. This approach is sensible because also the observational
constraint is taken to be a mean \Ae\ about which the values measured for
the single galaxies scatter. Still, in a more detailed analysis it would be
of great interest to resolve the SFHs of individual galaxies and to compare
the theoretical scatter in \Ae\ with the observed one.

In the following I will discuss uncertainties in chemical evolution that may
have an important impact on the results. Furthermore possible implications
on properties of cluster and field galaxies are mentioned.

\subsection{SN~Ia rate and stellar yields}
The model of chemical evolution applied for this analysis is calibrated such
that the basic local abundance features are covered \cite{TGB98}. Besides
stellar yields and IMF, this also includes the prescription of the Type Ia
supernova, which is adopted from Greggio \& Renzini \shortcite{GR83}.
According to this model, the maximum of the SN~Ia rate occurs roughly 1~Gyr
after the birth of the stellar population. It is obvious that this
time-scale has a great impact on the results presented here. From the solar
neighbourhood data, one cannot entirely exclude a SN~Ia rate which sets in
later and stronger. Such a model would basically delay (not prevent) the
decrease of the Mg/Fe ratio, hence in the hierarchical clustering scheme the
{\em global} populations of ellipticals would still not appear
$\alpha$-enhanced. However, the Mg/Fe ratios in the ISM would be higher {\em
at the major burst epochs}, leading to more \Ae\ in the burst populations.

Uncertainties in stellar yields \cite{TGB98} allow to raise the calculated
Mg/Fe ratios. Note, however, that the degree of \Ae\ used as the main
constraint in this paper ([$\alpha$/Fe]$=0.2$~dex) represents the lowest
limit determined from observational data. 

A very early chemical pre-enrichment by PopIII-like objects may induce
initial conditions in favour of high Mg/Fe ratios. The results in this
paper, however, are not seriously affected by such uncertainties, since it
is argued differentially to the solar neighbourhood chemistry on which the
chemical evolution model is calibrated. Thus details of the overall chemical
initial conditions in the early universe do not alter the conclusions.

\subsection{Selective mass loss}
Type~Ia supernovae explode later than Type~II, thus their ejection occurs at
different dynamical stages of the merging history. In a scenario in which
Type Ia products (basically iron) are lost easier, the accomplishment of
enhanced Mg/Fe ratios is simplified. The examination of this alternative,
however, requires a more detailed analysis of the possible outflow and star
formation processes during the merging history, which by far exceeds the
scope of this paper. On the other hand it is important to notice, that
$\alpha$-enhanced giant elliptical galaxies enrich -- for Salpeter IMF --
the intracluster medium (ICM) with Mg/Fe {\em underabundant} material
\cite{Th98b}. A selective loss mechanism as mentioned above further increases
this ICM-galaxy asymmetry \cite{Retal93} and therefore causes a more
striking disagreement with observational measurements which indicate solar
$\alpha$/Fe ratios in the ICM \cite{Metal96,IA97}.

\subsection{Cluster vs.\ Field}
The SFH discussed in this paper applies to {\em cluster} ellipticals. In
such a high density environment, the collapse of density peaks on galaxy
scale is boosted to high redshifts (K96). In the low-density surroundings of
the field (haloes of $10^{13}~\Msun$), instead, the situation is entirely
different. The K96 model predicts that objects in the field are not destined
to be ellipticals {\em or} spirals, but actually undergo transformations
between both types due to gas accretion (formation of a disc galaxy) or
merging of similar sized spirals (formation of an elliptical). As a
consequence, the objects we happen to observe as bright ellipticals in the
field should have mostly had a recent merger in the past 1.5 Gyr. This leads
to younger mean ages, but also to a more extended SFH. Figs.~\ref{fig1}
and~\ref{fig2} show that later bursts yield lower Mg/Fe ratios. From the
models, one should therefore expect a trend such that (bright) ellipticals
in the field are on average {\em less} $\alpha$-enhanced than their
counterparts in clusters (see also Thomas 1998a)\nocite{Th98a}. Thus, the
{\em intrinsic} difference between cluster and field ellipticals (K96),
which manifests itself most prominent in the properties of {\em bright}
elliptical galaxies, should be measurable in their $\alpha$-element
abundance patterns.

%=========================== C O N C L U S I O N ==========================

\section{Conclusion}
\label{conclusions}
In this paper, I analyse average star formation histories (SFH) as they
emerge from hierarchical clustering theory with respect to their capability
of producing $\alpha$-enhanced abundance ratios observed in elliptical
galaxies. For this purpose, the $V$-luminosity weighted age distributions
for the stellar populations of the model spiral and cluster elliptical
galaxy are adopted from Kauffmann (1996, K96)\nocite{K96}.

Owing to the constant level of star formation in spiral galaxies, their SFH
leads to roughly solar Mg/Fe ratios in the stars {\em and} in the ISM, in
agreement with observations in the Milky way. For elliptical galaxies,
hierarchical models predict more star formation at high redshift and
therefore significantly older mean ages. However, the calculations in this
paper show, that their SFH is still too extended in order to accomplish a
signifcant degree of {\em global} \Ae\ in the stars. Star bursts ignited by
the major merger when the elliptical forms, instead, provide a promising way
to produce metal-rich Mg/Fe overabundant populations, under the following
assumptions:
\begin{enumerate}
\item The nucleus predominantly consists of the burst population formed in
the major merger.
\item The burst population provides roughly 30 per cent of the total
$V$-light of the galaxy.
\item The IMF is significantly flattened ($x\la 0.8$) {\em during the burst} with
respect to the Salpeter value ($x=1.35$).
\item The efficiency of star formation during the burst has to be reduced to
roughly 50 per cent, in order to garuantee shallow metallicity gradients
within the galaxy.
\end{enumerate}
The burst populations and its proportions to the global populations turn out
to occupy a crucial role in producing Mg/Fe overabundance in the framework
of hierarchical galaxy fotmation. 

A direct consequence of the model is a slight gradient in \Ae\ in terms of
decreasing $\alpha$/Fe with increasing radius. A further implication is that
bright {\em field} ellipticals are predicted to exhibit lower $\alpha$/Fe
ratios than their {\em cluster} counterparts, due to the younger mean ages
and hence more extended SFHs of the former. A future theoretical task will
be to directly combine the chemical evolution of $\alpha$-elements and iron
with semi-analytic models in order to allow for a more quantitative analysis
than provided by this paper.

%========================= A C K N O W L E D G M E N T S =================

\section*{Acknowledgments}

I am very grateful to R.\ Bender and L.\ Greggio, who provided the mental
foundation stones of this work and gave important comments on the
manuscript. G.~Kauffmann, the referee of the paper, is particularly
acknowledged for very important suggestions that significantly improved the
first version. I also thank C.~Baugh, J.~Beuing, S.~Cole, N.~Drory,
H.~Kuntschner, C.~Lacey, C.~Maraston, D.~Mehlert and R.~Saglia for
interesting and helpful discussions. This work was supported by the
"Sonderforschungsbereich 375-95 f\"ur Astro-Teilchenphysik" of the Deutsche
Forschungsgemeinschaft.

% ========================= R E F E R E N C E S ===========================

%\bibliography{mnrasmnemonic,literature}
%\end{document}

\end{document}